\documentclass[12pt,journal,final,onecolumn]{IEEEtran}

\usepackage[final]{graphicx}
\usepackage{epsfig}
\usepackage[cmex10]{amsmath}
\usepackage{amssymb}
\usepackage{amsthm}
\usepackage{amsfonts}
\usepackage{bm}
\usepackage{cite}
\usepackage[tight,footnotesize]{subfigure}
\usepackage{hyperref} 
\usepackage{xcolor}

\graphicspath{{figs/}}

\input{niesen.def}

\newcommand{\scs}{\mathrel{:}}
\date{}

\begin{document}

\title{Energy-Efficient Communication over\\ the Unsynchronized Gaussian Diamond Network}
\author{Ritesh Kolte, Urs Niesen, and Piyush Gupta%
    \thanks{R.~Kolte is with the Department of Electrical Engineering at
    Stanford University. U.~Niesen and P.~Gupta are with Bell Labs,
    Alcatel-Lucent. Emails: rkolte@stanford.edu,
    urs.niesen@alcatel-lucent.com, piyush.gupta@alcatel-lucent.com}
    \thanks{This work was presented at the IEEE International Symposium on Information Theory 2014 in Honolulu, HI, USA.}
    \thanks{This work was supported in part by AFOSR under grant FA9550-09-1-0317.}%
}
\maketitle

\begin{abstract}
    Communication networks are often designed and analyzed  assuming
    tight synchronization among nodes. However, in applications that
    require communication in the energy-efficient regime of low
    signal-to-noise ratios, establishing tight synchronization among
    nodes in the network can result in a significant energy overhead.
    Motivated by a recent result showing that near-optimal energy
    efficiency can be achieved over the AWGN channel without requiring
    tight synchronization, we consider the question of whether the
    potential gains of cooperative communication can be achieved in the
    absence of synchronization. We focus on the symmetric Gaussian
    diamond network and establish that cooperative-communication gains
    are indeed feasible even with unsynchronized nodes. More precisely,
    we show that the capacity per unit energy of the unsynchronized
    symmetric Gaussian diamond network is within a constant factor of
    the capacity per unit energy of the corresponding synchronized network. To this end,
    we propose a distributed relaying scheme that does not require tight
    synchronization but nevertheless achieves most of the energy gains
    of coherent combining.
\end{abstract}

\section{Introduction} 

\subsection{Motivation}

An often implicit assumption in the design and analysis of communication
networks is that the clocks of different nodes in the network are
tightly synchronized. In practice, near-tight synchronization is
achieved via periodic transmission of pilot signals or via differential
modulation of data bits. These strategies result in little energy
overhead in the high-SNR regime, thus justifying the assumption of tight
synchronization. However, in applications such as space communication or
wireless sensor networks where energy efficiency is of critical
importance, communication has to take place in the low-SNR regime.  In
this regime, the energy overhead incurred in establishing near-tight
synchronization using the aforementioned approaches can be significant,
resulting in suboptimal energy efficiency. 

Recently, \cite{huang13} has established a fundamental limit on the
optimal energy efficiency, measured in bits per unit energy, that can be
achieved over the point-to-point AWGN channel without synchronization
between the transmitter and the receiver. This limit is nearly equal to
the optimal energy efficiency of the perfectly synchronized channel,
and, in contrast to the conventional approach, the corresponding
achievable scheme does not attempt to tightly synchronize clocks. 

Motivated by this result, we ask whether near-optimal energy efficiency
can be achieved in a distributed cooperative communication setting where
the participating nodes may not be synchronized. Multiple cooperating
nodes can potentially offer significant improvements in the
communication rate and energy efficiency by providing power and
multiplexing gains. However it is widely believed that establishing
tight synchronization is essential to achieve these gains.

As an example, consider a symmetric Gaussian multiple-access channel
where $K$ transmitters have a common message that is to be delivered to
the destination. With perfect synchronization, the transmitting nodes
can perform beamforming (i.e., achieve coherent combining of their
transmitted signals at the destination) thereby communicating $K$ times
more bits per unit energy than over a point-to-point channel. However,
if the nodes are not tightly synchronized, coherent combining of the
transmitted signals at the destination is not guaranteed.  On the other
hand, the performance of conventional schemes that first establish tight
synchronization can also be suboptimal \cite{huang13}. This poses the
question if these cooperative communication gains are still achievable
in the absence of node synchronization.

\subsection{Summary of Results}
\label{sec:intro_results}

In this paper, we consider the problem of communicating over an
unsynchronized symmetric $K$-relay Gaussian diamond network. To focus on
communicating in the energy-efficient regime, we adopt capacity per unit
energy as the performance metric instead of the usual notion of channel
capacity. Our main result is that the potential gains in energy
efficiency due to the presence of multiple relays are achievable up to a
constant multiplicative factor even in the absence of synchronization,
where the constant is independent of the channel gains and the number of
relays. 

To establish this result, we propose a distributed communication scheme
that achieves energy-efficient forwarding at the relays and coherent
combining of the signals at the destination by compensating for
synchronization errors.  This is done by concatenating an outer random
code with an inner repetition code. The inner code provides robustness
against the lack of synchronization and its length and scaling factor
are chosen so that every outer-code symbol can be transmitted reliably
with minimum energy.  The power levels of the outer random code at the
source and at the relays are chosen depending on the relative values of
the first-hop gain, the second-hop gain, and the number of relays in
order to obtain near-optimal energy efficiency.

\begin{figure}[htbp]
    \centering
    \subfigure[]{%
    \includegraphics{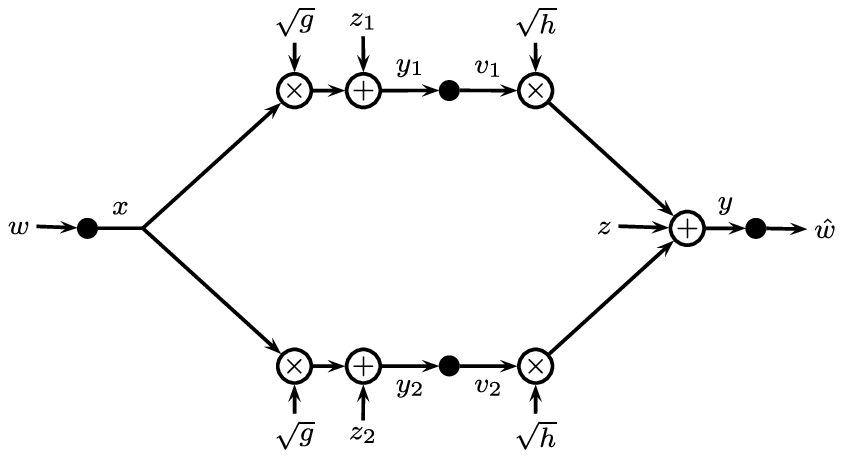}
    \label{fig:sync_diam}%
    }
    \hfill
    \subfigure[]{%
    \includegraphics{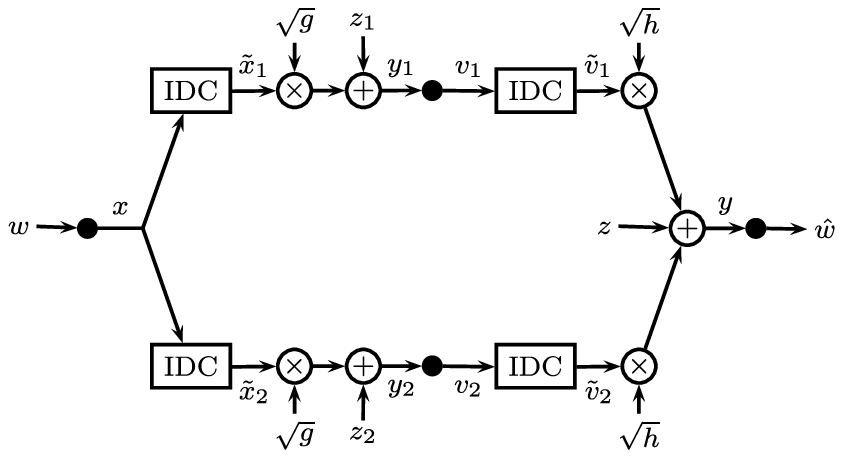}
    \label{fig:unsync_diam}%
    }
    \caption{Synchronized (a) and unsynchronized (b) two-relay Gaussian
    diamond networks.}
\end{figure}

Our model of the unsynchronized symmetric Gaussian diamond network is
depicted in Fig.~\ref{fig:unsync_diam} for $K = 2$ relays. The
unsynchronized network is obtained by introducing independent
insertion/deletion channels on all channels in the standard
(synchronized) symmetric Gaussian diamond network, depicted in
Fig.~\ref{fig:sync_diam}. The approach of modeling synchronization
errors via insertions/deletions is motivated by noting that if the clock
of a receiving node exhibits drift and jitter with respect to the clock
at the transmitting node, then the receiver may sample the transmitted
signal either faster than the transmitter, leading to insertions, or
slower, leading to deletions. 

\subsection{Related Work}

The insertion/deletion channel as a model for communication with
synchronization errors was introduced by Dobrushin in
\cite{dobrushin67}. Finding the capacity of this channel is still an
open problem, the main difficulty being the channel memory introduced by
the insertions and deletions. As a result of the difficulty in analyzing
the general insertion/deletion channel, most of the literature focuses
on a simpler special case, the binary deletion channel, for which good
approximations have been developed.  However, the exact capacity for
this special case too remains elusive.  We refer the reader to
\cite{mitzenmacher09} and references therein for a survey on this topic
up to 2009.  Some of the work that has appeared since 2009 is
\cite{kanoria10}, \cite{kalai10}, \cite{venkataramanan11},
\cite{drmota12}, \cite{iyengar11}. 

The notion of capacity per unit energy was analyzed for the synchronized
AWGN channel in \cite{golay49}. It is shown there that pulse-position
modulation achieves the capacity per unit energy of this channel.
Single-letter characterizations for the capacity per unit cost for
general cost functions and general synchronized discrete memoryless
channels were found in \cite{verdu90}. However, these results depend on
the channel being memoryless, whereas the channel considered here has
memory  due to the presence of insertions and deletions. Hence, these
results cannot be directly applied here. The pulse-position modulation
scheme was generalized for unsynchronized channels in~\cite{huang13},
where it was shown to achieve the capacity per unit energy of the
unsynchronized AWGN channel. We point out that, pulse-position
modulation itself cannot be extended to the unsynchronized diamond
network considered in this paper.  Indeed, unless the relays decode the
message sent by the transmitter, they end up spending a significant
amount of energy forwarding noise. However, requiring the relays to
decode the message can also be suboptimal---for instance, when the
source-relay channel is weak. 

Since its introduction in \cite{schein00}, there have been numerous
works analyzing the capacity of the synchronized Gaussian diamond
network. The achievable rates of two well-known relaying schemes,
decode-forward and amplify-forward, were analyzed for the two-relay
diamond network in \cite{schein00}. To counter the poor performance of
amplify-forward at low SNR, the bursty amplify-forward scheme was
proposed in \cite{schein01}. This scheme was shown to be approximately
optimal for the symmetric Gaussian diamond network with arbitrary number
$K$ of relays in \cite{niesen10b}, both in the sense of a uniform
additive gap of $1.8$ bits and a uniform multiplicative gap of a factor
$14$. 

The constant multiplicative approximation guarantee provided
in~\cite{niesen10b} implies that bursty amplify-and-forward also
achieves the capacity of the synchronized symmetric Gaussian diamond per
unit energy up to the same multiplicative gap. For the symmetric diamond
networks with only two relays, it was shown in \cite{parvaresh13} that
bursty amplify-forward and superposition-partial-decode-forward achieve
the capacity per unit energy up to a factor of at most $2.85$ and at
most $1.87$, respectively. The capacity per unit energy of the canonical
synchronized single-relay channel was approximated to within a factor
$1.7$ in~\cite{elgamal06}. A bursty amplify-and-forward scheme was also shown to achieve the optimal outage capacity per unit energy for the frequency-division relay channel at low SNR and at low outage probability \cite{avestimehr07}. It is worth pointing out that the bursty
constructions mentioned here do not generalize to the unsynchronized
diamond network considered in this paper, since, due to the
synchronization errors, the bursty codewords cannot be made to combine
constructively at the destination.

Different types of synchronization errors have been considered in the
literature. Insertion/deletion channels as used in this paper model
clock drift and jitter at the symbol level. A different approach is to
assume that the clocks at different nodes only differ by a constant
offset with respect to each other. For such errors, the offset is
typically either assumed to be equal to a multiple of the symbol
interval (frame asynchronism~\cite{massey72}) or equal to a length less
than one symbol interval (symbol asynchronism). The multiple-access
channel capacity region under symbol asynchronism and frame asynchronism has been analyzed in \cite{verdu89} and \cite{cover81,hui85}, respectively. Several recent works, \cite{chandar10} and \cite{shomorony12}, address the energy efficiency of bursty data communication. Here, asynchronism is not in the sense of symbol or frame asynchronism as described above, but in the sense that data arrives at the source sporadically at some random time unknown beforehand. 

Another model for asynchronism has been proposed and analyzed in \cite{yeung09}. Here, the asynchronism is modeled by stretching or shrinking the continuous-time transmitted signal by a time-dependent factor. This model can be thought of as the continuous-time analog of the asynchronism model that we use, however a number of additional assumptions are required to make the continuous-time problem mathematically tractable: the compress/stretch factor is assumed to lie between two positive numbers (i.e. bounded), the communication system is assumed to be noiseless and have infinite bandwidth, the allowed run-lengths lie in a closed positive interval and the input alphabet is finite.

\subsection{Organization}

We describe the problem setting formally in Section~\ref{sec:model}. The
main results are presented in Section~\ref{sec:main}. The proof of these
results are contained in Sections~\ref{sec:sync} and~\ref{sec:ach}.

\section{System Model}
\label{sec:model}

We first describe the behavior of the stand-alone insertion/deletion
channel (IDC) in Section~\ref{sec:model_IDC}. The model of the overall
diamond network is described in Section~\ref{sec:model_diamond}.

\subsection{Insertion/Deletion Channel}
\label{sec:model_IDC}

Let $(x[1], \dots, x[T])\in\mc{X}^T$ denote the input to the IDC. The
actions of the IDC are governed by an \iid sequence of states $(s[1],
\dots, s[T])\in\{0,1,2,\dots\}^T$. The state random variable $s[t]$
describes how many times input symbol $x[t]$ appears at the output of
the IDC. Let $(\tilde{x}[1], \dots,
\tilde{x}[\tilde{L}])\in\mc{X}^{\tilde{L}}$ denote the output of the
IDC, where $\tilde{L}$ is the random output length and is equal to
$\sum_{t=1}^T s[t].$ We refer the reader to \cite[Example 1]{huang13}
for an example of the operation of the IDC. We denote by
\begin{align*}
    \mu & \defeq \E(s[1]) \\
    \shortintertext{and} \\
    \sigma^2 & \defeq \var(s[1])
\end{align*}
the mean and the variance of the insertion/{\allowbreak}deletion
process, respectively, and we refer to any IDC with those parameters as
$\IDC(\mu, \sigma^2)$.  Here, $\mu$ and $\sigma^2$ can be interpreted as
capturing the drift and jitter of the receiver clock, respectively. As
noted in \cite{huang13}, in practice the parameter $\mu$ is close to
$1$, e.g., $\mu=1\pm 10^{-4}$.

\subsection{Unsynchronized Diamond Network}
\label{sec:model_diamond}

The unsynchronized diamond network is depicted in
Fig.~\ref{fig:unsync_diam} in Section~\ref{sec:intro_results}. The
source node transmits a message to the destination node with the help of
$K$ parallel relays. The channel inputs at time $t$ transmitted by the
source and $k$th relay are denoted by $x[t]$ and $v_k[t]$, respectively.
Each channel input passes through an independent IDC$(\mu,\sigma^2)$.
The output of the $k$th IDC on the first hop is denoted by
$\tilde{x}_k[\ell]$, while the output of the $k$th IDC on the second hop is
denoted by $\tilde{v}_k[\ell]$.  The channel outputs at time $\ell$ at the
destination and the $k$th relay are denoted by $y[\ell]$ and
$y_k[\ell]$, and are related to the IDC outputs as
\begin{align*}
    y_k[\ell] & \defeq \sqrt{g}\:\tilde{x}_k[\ell]+z_k[\ell], \\
    y[\ell] & \defeq \sqrt{h}\sum_{k=1}^K \tilde{v}_{k}[\ell]+z[\ell],
\end{align*}
where $\{z[\ell]\}_{\ell}, \{z_k[\ell]\}_{k,\ell}$ are independent and
identically distributed Gaussian random variables with mean zero and
variance one and independent of the corresponding channel inputs. The
channel gains $g$ and $h$ are assumed to be real positive numbers,
constant as a function of time and known throughout the network. Note
that due to the random insertions and deletions, the
lengths of the output sequence from different IDCs may not be
the same. However, in our achievable scheme, we consider only the
first 
\begin{equation*}
    L\defeq T\mu
\end{equation*}
symbols in the output sequence of any IDC, truncating or zero-padding as
necessary.\footnote{To simplify notation, we sometimes assume that
    quantities such as $\mu T$ are integers and omit $\floor{\cdot}$ and
    $\ceil{\cdot}$ operators.}

A $(T,M,P_1,P_2,\varepsilon)$ code for the unsynchronized diamond
network is a collection of functions
\begin{align*}
    f\from & \{1,\ldots, M\} \to \R^T, \\
    f_k\from & \R^{L} \to \R^T, \hspace{1.6cm} \forall k\in\{1,\ldots, K\},\\
    \phi\from & \R^{L}\to \{1,\ldots, M\},
\end{align*}
that satisfies the following properties.
\begin{itemize}
    \item The encoding function $f$ maps the message $w$,
        assumed to be uniformly distributed over the set $\{1,\ldots, M\}$, to
        the channel inputs (codeword)
        \begin{equation*}
            (x[t])_{t=1}^T\defeq f(w)
        \end{equation*}
        at the source node.
    \item The function $f_k$ maps the received channel outputs
        $(y_k[\ell])_{\ell =1}^{L}$ at the $k$th relay  to its transmitted codeword
        \begin{equation*}
            (v_k[t])_{t=1}^T\defeq f_k\big( (y_k[\ell])_{\ell =1}^{L} \big).
        \end{equation*}
        Note that the function $f_k$ is noncausal as described here.
        See Remark~\ref{rem:causal} for a discussion.
    \item The decoding function $\phi$ maps the received channel outputs
        $(y[\ell])_{\ell =1}^{L}$ at the destination node into a
        reconstruction 
        \begin{equation*}
            \hat{w} \defeq \phi\big( (y[\ell])_{\ell =1}^{L}\big)
        \end{equation*}
        of the message sent by the source node.
    \item Each codeword satisfies the power constraints
        \begin{align*}
            \frac{1}{T}\sum_{t=1}^T x^2[t] & \leq P_1, \\
            \frac{1}{T}\sum_{t=1}^T v_k^2[t] & \leq P_2, \ \quad\forall k\in\{1,\ldots, K\}.
        \end{align*}
    \item The average probability of error satisfies
        \begin{equation*}
            \Pp(\hat{w}\neq w)\leq\varepsilon.
        \end{equation*}
\end{itemize}

The total energy spent in transmitting the $\log M$ message bits by the
$(T,M,P_1,P_2,\varepsilon)$ code is at most $T(P_1+KP_2)$.
The \emph{rate per unit energy} achieved by this code is
\begin{equation*}
    \frac{\log M}{T(P_1+KP_2)},
\end{equation*}
where here and throughout this paper $\log(\cdot)$ and $\ln(\cdot)$
denote the logarithms to the bases $2$ and $e$, respectively.

\begin{definition}
    A rate of $\hat{R}$ bits per unit energy is \emph{achievable} if for
    every $\varepsilon>0$ and every large enough $M$ there exists a
    $(T,M,P_1,P_2,\varepsilon)$ code satisfying
    \begin{equation*}
        \frac{\log M}{T(P_1+KP_2)}\geq \hat{R}.
    \end{equation*}
    The \emph{capacity per unit energy} $\hat{C}$ is the supremum of
    achievable rates per unit energy.
\end{definition}

\begin{remark} 
    \label{rem:P1P2} 
    We emphasize that in the capacity per unit energy setting, the power
    constraints $P_1$ and $P_2$ are parameters to be optimized rather
    than fixed quantities as in the standard capacity setting. This is
    reflected by the code parameters being denoted by
    $(T,M,P_1,P_2,\varepsilon)$ instead of the usual $(T,M,\varepsilon)$.
\end{remark}

\begin{remark}
    \label{rem:causal} 
    To simplify the exposition, we are assuming noncausal encoding
    functions at the relays. This is not a restriction due to the
    layered nature of the network. The achievability scheme described in this paper can be adapted to the case of causal encoding functions by delaying the relay transmissions by a predetermined amount of time. This amount of time is chosen to be sufficiently large so that the relay encoding operations become causal. Note that this does not affect the
    number of bits transmitted per unit energy. A complicating issue is that, since the clocks of the
    relays are not synchronized with each other, the predetermined
    amount of time may not elapse simultaneously across the relays. As a
    result, they may start transmitting at different times. However, this mismatch can be handled using the same techniques as in the proof of Theorem~\ref{thm:unsync}. The resulting rate per unit energy with causal encoding functions is the same as with noncausal ones.
\end{remark}

\section{Main Results}
\label{sec:main}

The goal of this paper is to show that the capacity per unit energy of
the Gaussian diamond network is decreased by at most a constant factor
in the absence of synchronization. To this end, we start with an upper
bound on the capacity per unit energy of the \emph{synchronized}
Gaussian diamond network.

\begin{theorem}
    \label{thm:sync}
    The capacity per unit energy $\hat{C}_{\text{sync}}$ of the
    synchronized symmetric diamond network with $K$ relays and channel
    gains $g, h$ is upper bounded by
    \begin{equation*}
        \hat{C}_{\text{sync}} 
        \leq \frac{2}{\ln 2}\min\bigl\{Kg,\sqrt{Kgh},Kh\bigr\}.
    \end{equation*}
\end{theorem}

The proof of Theorem~\ref{thm:sync}, presented in Section
\ref{sec:sync}, builds on the capacity approximation for the symmetric
diamond relay network derived in~\cite{niesen10b}. The next result
states an achievable rate for the \emph{unsynchronized} Gaussian diamond
network.

\begin{theorem}
    \label{thm:unsync}
    The capacity per unit energy $\hat{C}_{\text{unsync}}$ of the
    unsynchronized symmetric Gaussian diamond network with $K$ relays,
    channel gains $g,h$, and average clock drift $\mu$ is lower bounded
    by
    \begin{equation*}
        \hat{C}_{\text{unsync}} 
        \geq \frac{\mu}{10}\min\bigl\{Kg,\sqrt{Kgh},Kh\bigr\}.
    \end{equation*}
\end{theorem}

The corresponding achievable scheme and the proof of this lower bound
are described in Section~\ref{sec:ach}. Note that the above result does not depend on the variances of the IDC. In fact, the IDCs need not have the same variance, and we only require that they be finite. Combining Theorems~\ref{thm:sync} and~\ref{thm:unsync}, we have the following corollary.

\begin{corollary}
    \label{thm:cor}
    The capacities per unit energy of the synchronized and
    unsynchronized symmetric Gaussian $K$-relay diamond networks
    satisfy
    \begin{equation*}
        \hat{C}_{\text{unsync}} 
        \geq \frac{\mu}{29}\hat{C}_{\text{sync}}.
    \end{equation*}
\end{corollary}

Corollary~\ref{thm:cor} states that the loss due to synchronization
errors for the Gaussian diamond network is at most a factor $29/\mu$.
As remarked earlier, $\mu\approx 1$ in practice. For those cases, the
loss due to synchronization errors is hence at most a factor of
approximately $29$. This is perhaps surprising, since tight
synchronization between nodes is often thought to be necessary to
achieve coherent combining of signals. The result here shows that this
tight synchronization is, in fact, not required to achieve essentially
the full energy-efficiency gains afforded by beamforming from the
relays.

Note that the IDC model used in this paper is physically meaningful  only when $\mu\approx 1$, i.e. when the clocks are not wildly out of synchronization. When $\mu \gg 1$, the unsynchronized channel model ``creates'' additional energy as compared to the synchronized channel, due to the availability of additional independently corrupted channel outputs at the receiver for each channel input. This effect increases the average received energy of the unsynchronized channel compared to the synchronized channel. This explains the fact that Corollary~\ref{thm:cor} implies that the lower bound on $\hat{C}_{\text{unsync}}$ becomes larger than $\hat{C}_{\text{sync}}$ for $\mu > 29.$ 
Thus, it is appropriate to compare $\hat{C}_{\text{unsync}}$ and $\hat{C}_{\text{sync}}$ only when $\mu\approx 1.$ For highly unsynchronized clocks, which is uncommon in practice, more sophisticated channel models need to be used. These models need to incorporate aspects of the modulating pulse, among other things, so that when multiple samples are taken, the SNR gets modified appropriately to avoid the artificial creation or destruction of energy.

As a consequence of the above results, we can identify three different
operating regimes for both the synchronized and unsynchronized diamond
network with respect to the parameters, $g$, $h$, and $K$.
\begin{itemize}
    \item $h < g/K$: In this regime, the gain $g$ of the source-relay
        links is much higher than the gain $h$ of the relay-destination
        links.  Under a path-loss model, this would correspond to the
        relays being located close to the source. By
        Theorems~\ref{thm:sync} and~\ref{thm:unsync}, the capacity per
        unit energy is proportional to $Kh$. 
        
        To understand the improvement provided by the relays, consider a
        point-to-point channel with a $K$-antenna source and a
        single-antenna destination with channel gains of magnitude $h$
        between them. The $K$ antennas at the source of this
        point-to-point channel can be used for transmit beamforming,
        resulting in a capacity per unit energy proportional to $Kh$.
        
        We thus see that in this regime the relays of the original
        diamond relay network can essentially be made to perform as if
        they were multiple antennas at the source.
        From~\cite{niesen10b}, we know that the minimum cut in this
        regime is the multiple-access cut separating the destination from the
        relays.

    \item $h \geq Kg$: In this regime, the gain $g$ of the source-relay
        links are much weaker than the gain $h$ of the relay-destination
        links. The relays can thus be thought of as being located
        close to the destination. By Theorems~\ref{thm:sync}
        and~\ref{thm:unsync}, the capacity per unit energy is
        proportional to $Kg$. 

        By a similar argument as above, we see that in this regime the
        relays in the network can essentially be made to perform as if
        they were multiple antennas at the receiver.
        From~\cite{niesen10b}, we know that the minimum cut in this
        regime is the broadcast cut separating the relays from the
        source.

    \item $g/K \leq h < Kg$: In this intermediate regime the
        source-relay gain $g$ is comparable to the relay-destination
        gain $h$. By Theorem~\ref{thm:sync} and~\ref{thm:unsync}, the
        capacity per unit energy is proportional to $\sqrt{Kgh}$.
        
        To get some insight into this behavior, consider the special
        case with equal channel gains $g=h$. The capacity per unit
        energy of the diamond network is then proportional to
        $\sqrt{Kgh}=\sqrt{K}g$. Comparing this to a point-to-point
        channel with gain $g$, we thus see that the relays in this
        regime provide a ``beamforming'' gain that scales as $\sqrt{K}$
        instead of $K$.  As identified in \cite{niesen10b}, the minimum
        cut in this case is neither the multiple-access cut nor the
        broadcast cut, but rather cuts that include only some of the
        relays.  
\end{itemize}

\begin{remark}\label{rem:unequal_mu}
We also note that by following the same steps as in the proof of Theorem~\ref{thm:unsync}, we can obtain a result similar to Corollary~\ref{thm:cor} for a model in which the means of the IDCs are not the same. Formally, if the $k$th IDC in the first hop has mean $\tilde{\mu}_{1,k}$ and the $k$th IDC in the second hop has mean $\tilde{\mu}_{2,k}$, for $k\in\{1,2,\dots , K\}$, then we have
\begin{equation}\label{eq:unequal_mu}\hat{C}_{\text{unsync}} 
        \geq \frac{\tilde{\mu}}{29}\hat{C}_{\text{sync}},\end{equation}
        where $$\tilde{\mu}\triangleq 2\left(\frac{1}{\min_{1\leq k\leq K} \tilde{\mu}_{1,k}} + \frac{1}{K}\sum_{k=1}^{K}\frac{1}{\tilde{\mu}_{2,k}}\right)^{-1}.$$
        Elaborating arguments for this case are presented in Section~\ref{subsec:unequal_mu}.
\end{remark}

\section{Proof of Theorem~\ref{thm:sync}}
\label{sec:sync}

We start by recalling a result from \cite[Theorem~2]{niesen10b} giving
an upper bound on the capacity of the \emph{synchronized} diamond
network.

\begin{theorem}
    \label{thm:cap_up}
    For every symmetric diamond network with $K\geq 2$ relays, channel
    gains $g,h$ and power constraints $P_1,P_2$, the capacity $C(P_1,
    P_2)$ is upper bounded by 
    \begin{equation*}
        \begin{cases}
            \tfrac{1}{2}\log\big(1+K\min\{P_1g,KP_2h\}\big),
            & \text{if $\max\{P_1g,KP_2h\} \geq 1$} \\
            \tfrac{1}{2}\log(1+KP_1g),
            & \text{if $\max\{P_1g,KP_2h\} < 1$, $P_1g \leq P_2h$} \\
            \tfrac{1}{2}\log\big(1+2K^2P_1P_2gh\big) + \frac{1}{2},
            & \text{if $\max\{P_1g,KP_2h\} < 1$, $P_1g \in(P_2h,K^2P_2h)$, $K\sqrt{P_1P_2gh}\geq 1$} \\
            \log(1+2K\sqrt{P_1P_2gh}),
            & \text{if $\max\{P_1g,KP_2h\} < 1$, $P_1g \in(P_2h,K^2P_2h)$, $K\sqrt{P_1P_2gh} < 1$} \\
            \tfrac{1}{2}\log(1+K^2P_2h),
            & \text{if $\max\{P_1g,KP_2h\} < 1$, $P_1g \geq K^2P_2h$}.
        \end{cases}
    \end{equation*}
\end{theorem}

\begin{figure}[t]
    \centering
    \includegraphics{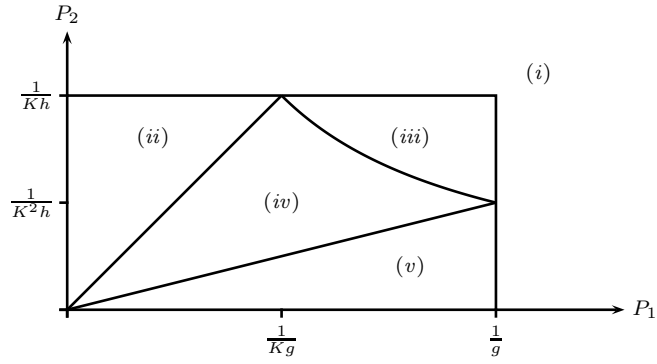}
    \caption{Different SNR regimes in Theorem~\ref{thm:cap_up}}
    \label{fig:snr_reg}
\end{figure}

Of the five regimes in Theorem~\ref{thm:cap_up}, cases $(i)$ and $(iii)$
can be considered as the high-SNR regimes, while $(ii)$, $(iv)$ and
$(v)$ can be considered as the low-SNR regimes (see
Fig.~\ref{fig:snr_reg}). We start by arguing that we can restrict
attention to only the low-SNR regimes.

For power constraints $P_1, P_2$, denote by $C(P_1,P_2)$ the
corresponding capacity. The rate per unit energy achievable for fixed
powers $(P_1,P_2)$ is then 
\begin{equation*}
    \frac{C(P_1,P_2)}{P_1+KP_2},
\end{equation*}
and the capacity per unit energy is
\begin{equation}
    \label{eq:optimization}
    \sup_{P_1, P_2 \geq 0} \frac{C(P_1,P_2)}{P_1+KP_2}.
\end{equation}

We argue that reducing both power constraints by a factor
$\lambda\in(0,1)$ cannot decrease this rate per unit energy.  Fix such a
value of $\lambda$ as well as transmit powers $P_1, P_2$.  Consider
communicating during only a fraction $\lambda$ of the total time using
the optimal communication scheme for powers $P_1$ and $P_2$.  The rate
of communication is $\lambda C(P_1,P_2)$, and the average power
constraints are $\lambda P_1$ and $\lambda P_2$.  Hence, the achieved
rate per unit energy is
\begin{equation*}
    \frac{\lambda C(P_1,P_2)}{\lambda P_1+K\lambda P_2} 
    = \frac{C(P_1,P_2)}{P_1+KP_2}.
\end{equation*}
Since this is only one particular strategy for communicating with
average powers $(\lambda P_1, \lambda P_2)$, we have
\begin{equation*}
    \lambda C(P_1, P_2) \leq C(\lambda P_1, \lambda P_2), 
\end{equation*}
which implies that
\begin{equation*}
    \frac{\lambda C(P_1,P_2)}{\lambda P_1+K\lambda P_2} 
    \leq \frac{ C(\lambda P_1,\lambda P_2)}{\lambda P_1+K\lambda P_2}.
\end{equation*}
Combining this, we obtain 
\begin{equation*}
    \frac{C(P_1,P_2)}{P_1+KP_2}
    \leq \frac{C(\lambda P_1,\lambda P_2)}{\lambda P_1+K\lambda P_2}.
\end{equation*}

This shows that the capacity per unit energy cannot decrease if $P_1$
and $P_2$ are replaced by $\lambda P_1$ and $\lambda P_2$, respectively,
for any $0<\lambda < 1$. Hence, we can restrict the optimization over
transmit powers in~\eqref{eq:optimization} to any arbitrarily small
neighborhood of the origin.  This allows us to ignore the first and
third regimes in Theorem~\ref{thm:cap_up} (see again
Fig.~\ref{fig:snr_reg}).

We now derive an upper bound on the capacity per unit energy in each of
the three low-SNR regimes $(ii)$, $(iv)$, and $(v)$. For ease of
notation, we do not explicitly state the nonnegativity condition on
$P_1,P_2$ in the following analysis. For the same reason, we also do not
state explicitly the condition $\max\{P_1g,KP_2h\}<1$, which is common
to all the low-SNR regimes, and assume implicitly that $P_1$ and $P_2$
are small enough to satisfy this condition.

\begin{itemize}
    \item Assume first that $(P_1,P_2)$ are chosen such that we operate
        in regime $(iv)$, i.e., $P_2h<P_1g<K^2P_2h$ and
        $K\sqrt{P_1P_2gh} < 1$. The achievable rate per unit energy in
        this regime is upper bounded by (not mentioning the condition
        $K\sqrt{P_1P_2gh} < 1$ explicitly)
        \begin{align*}
            \sup_{(P_1,P_2):\: P_2h<P_1g<K^2P_2h} 
            & \: \frac{\log(1+2K\sqrt{P_1P_2gh})}{P_1+KP_2} \\
            & \stackrel{(a)}{\leq} \sup_{(P_1,P_2):\: P_2h<P_1g<K^2P_2h}
            \: \frac{1}{\ln 2}\cdot\frac{2K\sqrt{P_1P_2gh}}{P_1+KP_2} \\
            & \stackrel{(b)}{=} \sup_{x:\: h/g\leq x\leq K^2h/g}
            \: \frac{2K\sqrt{gh}}{\ln 2}\cdot \frac{\sqrt{x}}{x+K} ,
        \end{align*}
        where $(a)$ follows from $\ln(1+x)\leq x$ and $(b)$ by making
        the change of variable $x = P_1/P_2$. Now, for $x\geq 0$, 
        \begin{equation*}
            \frac{\sqrt{x}}{x+K}
            \leq \min\Bigl\{ \frac{1}{\sqrt{x}},
            \frac{\sqrt{x}}{\max\{x,K\}}, 
            \frac{\sqrt{x}}{K} \Bigr\}, 
        \end{equation*}
        so that
        \begin{align*}
            \sup_{x:\: h/g\leq x\leq K^2h/g} \: \frac{\sqrt{x}}{x+K} 
            & \leq \sup_{x:\: h/g\leq x\leq K^2h/g} 
            \min\Bigl\{ \frac{1}{\sqrt{x}},
            \frac{\sqrt{x}}{\max\{x,K\}}, 
            \frac{\sqrt{x}}{K} \Bigr\} \\
            & \leq \min\Bigl\{ \frac{\sqrt{g}}{\sqrt{h}},
            \frac{1}{\sqrt{K}}, 
            \frac{\sqrt{h}}{\sqrt{g}} \Bigr\}.
        \end{align*}
        We thus obtain 
        \begin{align*}
            \sup_{(P_1,P_2):\: P_2h<P_1g<K^2P_2h} 
            \: \frac{\log(1+2K\sqrt{P_1P_2gh})}{P_1+KP_2}
            \leq \frac{2}{\ln 2}\min\bigl\{Kg,\sqrt{Kgh},Kh\bigr\}.
        \end{align*}
    \item Assume next that $(P_1,P_2)$ are chosen such that we operate
        in regime $(ii)$, i.e., $P_1g \leq P_2h$. The achievable rate
        per unit energy in this regime is upper bounded by
        \begin{align*}
            \sup_{(P_1,P_2):\: P_1g \leq P_2h} 
            & \: \frac{\frac{1}{2}\log(1+KP_1g)}{P_1+KP_2} \\
            & = \sup_{P_1}\:\sup_{P_2\geq P_1g/h} 
            \: \frac{\frac{1}{2}\log(1+KP_1g)}{P_1+KP_2}\\
            & \stackrel{(a)}{=} \frac{1}{2\ln 2}\cdot \frac{Kg}{1+Kg/h} \\
            & \stackrel{(b)}{\leq} \frac{2}{\ln 2}\min\bigl\{Kg,\sqrt{Kgh},Kh\bigr\},
        \end{align*}
        where $(a)$ follows by setting $P_2 = P_1g/h$ and letting
        $P_1\to 0$, and where $(b)$ follows from
        \begin{equation}
            \label{eq:amgm}
            \frac{Kg}{1+Kg/h} 
            \leq \sqrt{Kgh}/2, 
        \end{equation}
        which can be proved using the arithmetic-mean--geometric-mean
        inequality.
    \item Assume finally that $(P_1,P_2)$ are chosen such that we
        operate in regime $(v)$, i.e., $P_1g \geq K^2P_2h$.  The
        achievable rate per unit energy in this regime is upper bounded
        by
        \begin{align*}
            \sup_{(P_1,P_2):\: P_1g \geq K^2P_2h} 
            & \: \frac{\frac{1}{2}\log(1+K^2P_2h)}{P_1+KP_2} \\
            & = \sup_{P_2} \:\sup_{P_1\geq K^2P_2h/g} 
            \: \frac{\frac{1}{2}\log(1+K^2P_2h)}{P_1+KP_2}\\
            & \stackrel{(a)}{=} \frac{1}{2\ln 2}\cdot \frac{Kh}{1+Kh/g} \\
            & \stackrel{(b)}{\leq} \frac{2}{\ln 2}\min\bigl\{Kg,\sqrt{Kgh},Kh\bigr\},
        \end{align*}
        where $(a)$ follows by setting $P_1 = K^2P_2h/g$ and letting
        $P_2\to 0$, and where $(b)$ follows again by~\eqref{eq:amgm}.
\end{itemize}

The achievable rate per unit energy is thus upper bounded by
\begin{equation*}
    \frac{2}{\ln 2}\min\bigl\{Kg,\sqrt{Kgh},Kh\bigr\},
\end{equation*}
in all three low-SNR regimes $(ii)$, $(iv)$, and $(v)$, 
concluding the proof of Theorem~\ref{thm:sync}.\hfill\IEEEQED

\section{Proof of Theorem~\ref{thm:unsync}}
\label{sec:ach}

The basic construction of the achievable scheme is as follows. The
transmitters employ a codebook in which each codeword consists of a
series of long and low-power pulses. This codebook can be described as
the concatenation of an outer random code with an inner repetition code.
The length of the pulses ensures robustness of the codewords with
respect to insertions/deletions and the low power ensures energy
efficiency of the communication. In order to achieve low error
probability and high energy efficiency, every receiving node needs to be
reasonably certain about the location of the different pulses in the
transmitted signal and there should be coherent combining of the relay
transmissions at the destination. We meet these two requirements by
choosing the code parameters appropriately.

\begin{figure}[!tp]
    \centering
    \includegraphics{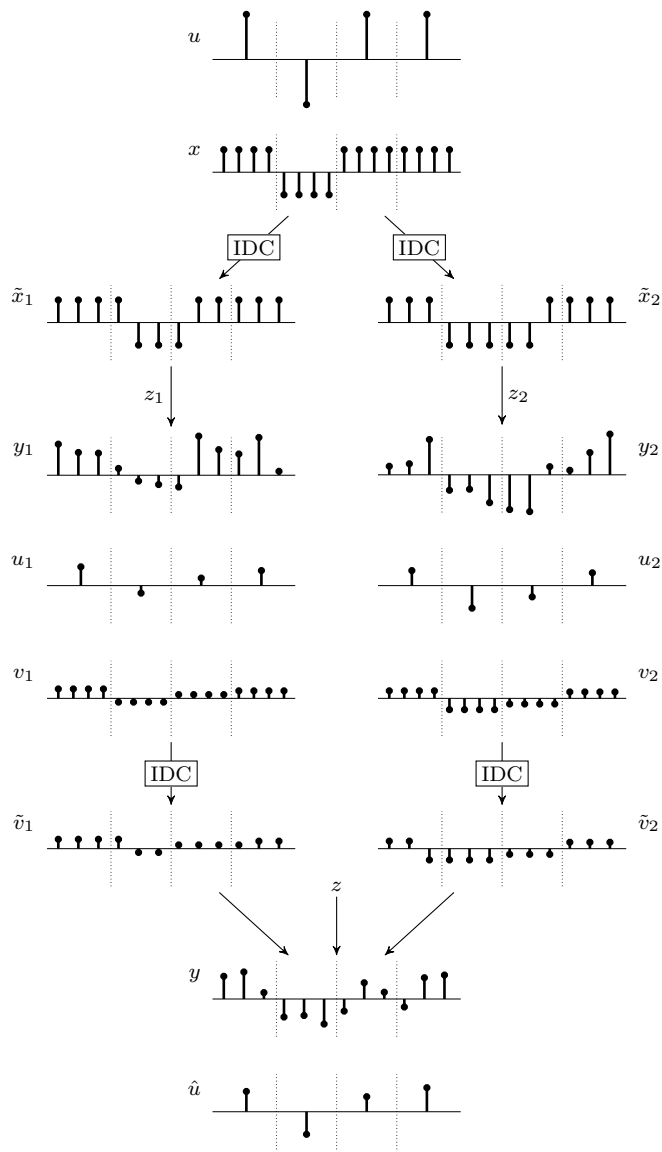}
    \caption{An illustration of the achievable scheme for a two-relay
        diamond network with IDC parameter $\mu=3/4$. The notation is
        the same as in Fig.~\ref{fig:unsync_diam} in
        Section~\ref{sec:intro_results}.  In the figure, we use an outer
        code of block length $N=4$ (see signals $u$, $u_k$, and
        $\hat{u}$) together with an inner repetition code of block
        length $N' = 4$ (see signals $x$ and $v_k$). The block length of
        the combined code is $T=NN'=16$.  Each relay forms an estimate
        of the symbols in the outer code by averaging every $\mu N' = 3$
        consecutive channel outputs (see the transformation from $y_k$
        to $u_k$). This estimate is then re-encoded using the same inner
        repetition code (see the transformation from $u_k$ to $v_k$).
        The destination uses the same averaging to compute an estimate
        $\hat{u}$ of the symbols in the outer code. The decoding of the
        outer code itself is done using the maximum-likelihood sequence
        detector. Note that only the destination, but \emph{not} the
        relays, decode the outer code.}
    \label{fig:notation_illus}
\end{figure}

Encoding at the source, processing at the relays, and decoding at the
destination are illustrated in Fig.~\ref{fig:notation_illus}.  We
provide a detailed description of the scheme in
Section~\ref{sec:ach_construction}. The analysis of this scheme is
contained in Section~\ref{sec:ach_analysis}.  We point out that we make
conservative choices for the block length to keep the analysis simple.
The block length can be made significantly smaller than presented in the
following analysis at the cost of a more complicated analysis.

\subsection{Construction}
\label{sec:ach_construction}

\subsubsection*{Encoding at the Source} 

The source first constructs an outer code by randomly and independently
generating $(u[n])_{n=1}^N$ for each message $w\in\{1,2,\dots, M\}$
according to the uniform distribution over the set
$\{-\sqrt{P_1},\sqrt{P_1}\}^N$. For simplicity of notation, we do not
explicitly indicate the dependence of the sequence $(u[n])_{n=1}^N$ on
the message $w$. The energy of each of these sequences is $NP_1$. To
form the transmission sequences, the source then employs an inner code.

This inner code is a simple repetition code, attenuated so that the total
energy remains unchanged. Specifically, the channel inputs
$(x[t])_{t=1}^T$ with 
\begin{equation*}
    T \defeq N^5
\end{equation*}
are formed by
\begin{itemize}
    \item repeating $N'$ times each symbol in $(u[n])_{n=1}^N$,
        where 
        \begin{equation}
            \label{eq:nprime}
            N'\defeq N^4, 
        \end{equation}
        and
    \item attenuating by a factor $1/\sqrt{N'}$.
\end{itemize}
As a result, the channel inputs are piecewise constant sequences
containing $N$ pieces, each piece having length equal to $N'$, and
height either $+\sqrt{P_1/N'}$ or $-\sqrt{P_1/N'}$.

Formally, the $t$th channel input $x[t]$ is given by 
\begin{equation*}
    x[t] \defeq 
    \frac{1}{\sqrt{N'}}u\left[\:\left\lfloor\frac{t-1}{N'}\right\rfloor+1\:\right]
\end{equation*}
for all $t\in\{1,2,\dots, T\}$. The length of each channel input 
is $T = NN' = N^5$, and the total energy used at the source is
\begin{equation}
    \label{eq:energy1}
    NN'\cdot\frac{1}{N'}\cdot P_1 = NP_1. 
\end{equation}

\subsubsection*{Processing at the Relays}

Relay $k$ divides its received signal $(y_k[\ell])_{\ell=1}^L$ with $L=
\mu T$ into $N$ consecutive blocks of length $\mu N'$. It forms a
normalized sum of the received signal in each block resulting in the
sequence $(u_k[n])_{n=1}^N$ of length $N$. Formally, the $n$th symbol
$u_k[n]$ is given by
\begin{equation*}
    u_k[n] 
    \defeq \frac{1}{\sqrt{\mu N'}}
    \sum_{\ell = (n-1)\mu N' + 1}^{n\mu N'} y_k[\ell]
\end{equation*}
for all $n\in\{1,2,\dots, N\}$. 

Relay $k$ then forms its own transmit sequence $(v_k[t])_{t=1}^T$  using
the attenuated repetition code, except that instead of attenuating by a
factor $1/\sqrt{N'}$, it uses the attenuation factor $\alpha/\sqrt{N'}$,
where 
\begin{equation}
    \label{eq:alpha}
    \alpha\defeq \sqrt{\frac{P_2}{1+\mu gP_1}}.
\end{equation}
Formally,
\begin{equation*}
    v_k[t] \defeq 
    \frac{\alpha}{\sqrt{N'}}u_k\left[\:\left\lfloor\frac{t-1}{N'}\right\rfloor + 1\:\right]
\end{equation*}
for all $t\in\{1,2,\dots, T\}$.

The amplification factor is chosen to limit the energy spent at the
relay. Indeed, the total energy spent by each relay is no more than 
\begin{equation}
    \label{eq:energy2}
    NN' \cdot \frac{\alpha^2}{N'} \cdot (\mu g P_1+1)
    = NP_2,
\end{equation}
where the $+1$ term accounts for the forwarded receiver noise.

\subsubsection*{Decoding at the Destination}

The destination performs a similar operation as the relays. It divides
its received signal $(y[\ell])_{\ell=1}^L$ into $N$ consecutive blocks
of length $\mu N'$. It forms the normalized sum of the received signal in
each block resulting in the sequence $(\hat{u}[n])_{n=1}^N$ of length
$N$. Formally,
\begin{equation*}
    \hat{u}[n] \defeq \frac{1}{\sqrt{\mu N'}}
    \sum_{\ell = (n-1)\mu N' + 1}^{n\mu N'} y[\ell]
\end{equation*}
for all $n\in\{1,2,\dots, N\}$.  

The destination considers the end-to-end channel from $(u[n])_{n=1}^N$
to $(\hat{u}[n])_{n=1}^N$ established as a result of the operations
described above and attempts to decode the transmitted message using
maximum-likelihood decoding.

\subsection{Analysis}
\label{sec:ach_analysis}
The analysis of the probability of error consists of the following key
steps.

\begin{itemize}
    \item The randomness of the IDCs causes the constant pieces in the
        first-hop-IDC outputs $\tilde{x}_k$ to not align perfectly with
        the blocks that the relays divide their received signal into,
        e.g., the constant pieces in $\tilde{x}_k$ in
        Fig.~\ref{fig:notation_illus} do not align with the dotted
        vertical lines. This manifests as inter-symbol interference (ISI)
        in the signal $u_k$. The first step is to
        show that this ISI is negligible as $N\rightarrow\infty$.
    \item For high energy efficiency, it is crucial to achieve
        a beamforming gain at the destination. To this end, we
        argue that, as $N\to\infty$, the $n$th piece in every
        second-hop IDC output $\tilde{v}_k$ approximately align at the
        destination.
    \item To operate within a constant factor of the optimal energy
        efficiency, the values of $P_1$ and $P_2$ need to be chosen
        appropriately depending on the system parameters $g,h,K$.
    \item Finally, the restriction to the binary alphabet
        $\{-\sqrt{P_1},\sqrt{P_1}\}$ (which simplifies the analysis of
        ISI) results in a rate loss, which needs to be bounded.
\end{itemize}

Fix a target error probability $\varepsilon >0$. Denote by $\mc{E}$ the
decoding error event, and let $\mc{E}_{\text{IDC}}$ denote the event
that any piece in the output of any IDC behaves far from expectation
(defined formally below). The probability of error
$\Pp(\mc{E})$ can be upper bounded as
\begin{align}
    \label{eq:error}
    \Pp(\mc{E}) 
    & = \Pp(\mc{E}\mid\mc{E}_{\text{IDC}})\Pp(\mc{E}_{\text{IDC}}) 
    + \Pp(\mc{E}\mid\mc{E}_{\text{IDC}}^c)\Pp(\mc{E}_{\text{IDC}}^c) \nonumber\\
    & \leq \Pp(\mc{E}_{\text{IDC}})+ \Pp(\mc{E}\mid\mc{E}_{\text{IDC}}^c)
\end{align}
We next argue that, for large enough $N$, each of these two terms can be
made less than $\varepsilon/2$.

\subsection*{Behavior of the IDCs}

We start with the analysis of the term $\Pp(\mc{E}_{\text{IDC}})$
in~\eqref{eq:error}. To this end, we establish that the behavior of
every IDC in the network concentrates around its expectation.

Let $(s_k[t])_{t=1}^T$ denote the state sequence of the $k$th IDC. The
input sequence to any IDC is a piecewise constant sequence containing
$N$ pieces. Each piece has length $N'$ and the $n$th piece starts at
position $(n-1)N'+1$ and ends at position $nN'$. The output of the IDC
is also a piecewise constant sequence, but the pieces now have random
lengths with expected value $N'\mu$.

Consider the $n$th piece in the output of the $k$th IDC. Let
$\mc{E}_{k,n}$ denote the event that this piece starts outside the
interval 
\begin{equation*}
    {((n-1)\mu N' + 1-\nu \scs (n-1)\mu N' + 1+\nu)}
\end{equation*}
or has length outside the interval 
\begin{equation*}
    (\mu N'-\beta \scs \mu N'+\beta), 
\end{equation*}
where 
\begin{align}
    \label{eq:nudef}
    \nu & \defeq N^{7/2} \\ 
    \shortintertext{and} \nonumber\\
    \label{eq:betadef}
    \beta & \defeq N^{3}.
\end{align}
Here and in the remainder of this paper we use the notation $[a \scs b]$
to denote all integers in the closed interval $[a,b]$, i.e.  $[a \scs b]
= [a,b]\cap \Z$, and similar for $(a \scs b)$.  Denote by 
\begin{equation*}
    \mc{E}_{\text{IDC}}
    \defeq \cup_{k=1}^{2K}\cup_{n=1}^N \mc{E}_{k,n}
\end{equation*}
the event that any piece in the output of any IDC behaves far from
expectation

The probability that the $n$th piece in the output of the $k$th IDC
starts outside the interval 
\begin{equation*}
    {((n-1)\mu N' + 1-\nu\scs (n-1)\mu N' + 1+\nu)}
\end{equation*}
is equal to
\begin{align*}
    \Pp\biggl(\biggl| \sum_{t=1}^{(n-1)N'} s_k[t] -
    (n-1)\mu N'\biggr|\geq \nu\biggr) 
    & \stackrel{(a)}{\leq} \frac{(n-1)N'\sigma^2}{\nu^2} \\
    & \leq \frac{NN'\sigma^2}{\nu^2} \\
    & \stackrel{(b)} = \frac{N^5\sigma^2}{(N^{7/2})^2} \\
    & = \frac{\sigma^2}{N^2},\end{align*}
where $(a)$ follows from Chebyshev's inequality and the definition of
$\mu$ and $\sigma^2$ as the mean and variance of $s_k[t]$, and where
$(b)$ follows from the definition of $N'$ in~\eqref{eq:nprime}.

The probability that the $n$th piece in the output of the $k$th IDC has
length outside 
\begin{equation*}
    (\mu N'-\beta\scs \mu N'+\beta)
\end{equation*}
is equal to
\begin{align*}
    \Pp\biggl(\biggl|\sum_{t=(n-1)N'+1}^{nN'} s_i[t] -
    \mu N'\biggr|\geq \beta\biggr)
    & \stackrel{(a)}{\leq} \frac{N'\sigma^2}{\beta^2} \\
    & = \frac{N^4\sigma^2}{(N^3)^2}  \\
    & = \frac{\sigma^2}{N^2},
\end{align*}
where $(a)$ follows again from Chebyshev's inequality.

Hence, by the union bound, 
\begin{equation*}
    \Pp(\mc{E}_{k,n})
    \leq \frac{2\sigma^2}{N^2}.
\end{equation*}
By the union bound again, we thus have
\begin{align}
    \label{eq:error1}
    \Pp\left(\mc{E}_{\text{IDC}}\right) 
    & = \Pp\left(\cup_{k=1}^{2K}\cup_{n=1}^N \mc{E}_{k,n}\right) \nonumber\\
    & \leq \sum_{k=1}^{2K}\sum_{n=1}^N \Pp(\mc{E}_{k,n}) \nonumber\\
    & \leq \frac{4K\sigma^2}{N} \nonumber\\
    & \leq \varepsilon/2
\end{align}
for $N$ large enough.

\subsection*{Analysis of the Received SNR on the End-to-End Channel}

We continue with the analysis of the term
$\Pp(\mc{E}\mid\mc{E}_{\text{IDC}}^c)$ in~\eqref{eq:error}.  We will
argue that this term is also upper bounded by $\varepsilon/2$ for
rate per unit energy less than or equal to 
\begin{equation*}
    \frac{\mu}{10}\min\bigl\{Kg,\sqrt{Kgh},Kh\bigr\}
\end{equation*}
and $N$ large enough.  For this purpose, we provide a worst-case lower
bound on the received SNR of every symbol of the outer code. This allows
us to lower bound the achievable rate, which in turn leads to the
aforementioned lower bound on the capacity per unit energy.

\subsubsection*{First Hop}

For the analysis of the first hop, number the IDCs from the source to
the relays as $1$ through $K$. Consider the operation of relay $k$ on
the received signal.  Without loss of generality, assume that the $n$th
symbol in the outer code at the source is 
\begin{equation*}
    u[n] = +\sqrt{P_1}.
\end{equation*}
The relay sums up its received signal in the block 
\begin{equation*}
    [(n-1)\mu N' + 1 \scs n\mu N'] 
\end{equation*}
to form $u_k[n]$, which is treated as the noise-corrupted signal
corresponding to the transmitted symbol $u[n]$.\footnote{To maintain a
    clear and consistent terminology, we use ``piece'' to denote an
    interval over which the signal is constant, and ``block'' to denote
    an interval of the form $[(n-1)N' + 1 \scs nN']$ or $[(n-1)\mu N' + 1
    \scs n\mu N']$, depending on whether it is the input or output of an
    IDC, respectively.  To differentiate between the signal component
    from the noise component in the received signal, we use ``signal''
    to denote only the component due to source transmission, and use
    ``received signal'' to denote signal plus any noise component.  }
\begin{align*}
    u_k[n] 
    & = \frac{1}{\sqrt{\mu N'}}\sum_{\ell = (n-1)\mu N' + 1}^{n\mu N'} y_k[\ell] \\
    & = \frac{1}{\sqrt{\mu N'}}\sum_{\ell = (n-1)\mu N' + 1}^{n\mu N'}
    \bigl(\sqrt{g}\:\tilde{x}_k[\ell] + z_k[\ell]\bigr)
\end{align*}
Using $\hat{z}_k[n]$ to denote the quantity 
\begin{equation*}
    \hat{z}_k[n]
    \defeq
    \frac{1}{\sqrt{\mu N'}}\sum_{\ell = (n-1)\mu N' + 1}^{n\mu N'} z_k[\ell]
    \sim\mc{N}(0,1),
\end{equation*}
we can rewrite the expression for $u_k[n]$ as
\begin{equation}
    \label{eq:op_IDC}
    u_k[n] 
    = \biggl(\sqrt{\frac{g}{\mu N'}}\sum_{\ell = (n-1)\mu N' + 1}^{n\mu N'}
    \tilde{x}_k[\ell]\biggr) + \hat{z}_k[n].
\end{equation}
Similarly,
\begin{align}
    \label{eq:op_IDC2}
    u_k[n-1] & = \biggl(\sqrt{\frac{g}{\mu N'}}\sum_{\ell = (n-2)\mu N' + 1}^{(n-1)\mu N'}
    \tilde{x}_k[\ell]\biggr) + \hat{z}_k[n-1], \\
    \label{eq:op_IDC3}
    u_k[n+1] & = \biggl(\sqrt{\frac{g}{\mu N'}}\sum_{\ell = n\mu N' + 1}^{(n+1)\mu N'}
    \tilde{x}_k[\ell]\biggr) + \hat{z}_k[n+1]. 
\end{align}

The input to any IDC is a piecewise constant signal where the pieces
\begin{equation*}
    [(n-1)N'+1 \scs nN']
\end{equation*}
align precisely with the blocks 
\begin{equation*}
    [(n-1)N'+1 \scs nN']. 
\end{equation*}
However, due to the randomness of the IDC, the $n$th piece in the output
of the IDC may not align with the $n$th output block
\begin{equation*}
    [(n-1)\mu N' +1\scs n\mu N'].
\end{equation*}
This block can contain signal resulting from neighboring pieces in the
input sequence, which correspond to $u[n-1]$ and $u[n+1]$. Since relay
$k$ forms $u_k[n]$ according to \eqref{eq:op_IDC}, this leakage from
neighboring symbols manifests itself as inter-symbol interference.
However, conditioned on $\mc{E}_{\text{IDC}}^c$, we expect that the
block 
\begin{equation*}
    [(n-1)\mu N' +1 \scs n\mu N']
\end{equation*}
contains a significant portion of the $n$th piece so that the
inter-symbol interference is low. This is indeed the case as formalized
below.

For the worst inter-symbol interference, we assume that the neighboring
symbols $u[n-1]$ and $u[n+1]$ are both equal to $-u[n] = - \sqrt{P_1}$.
Conditioned on $\mc{E}_{\text{IDC}}^c$, we know that the $n$th piece in
the output of the IDC starts in the interval 
\begin{equation*}
    ((n-1)\mu N' + 1-\nu\scs (n-1)\mu N' + 1+\nu)
\end{equation*}
and has length in the interval 
\begin{equation*}
    (\mu N'-\beta\scs \mu N'+\beta).
\end{equation*}
Hence the block 
\begin{equation*}
    [(n-1)\mu N' + 1\scs n\mu N']
\end{equation*}
in the output of the IDC contains a constant piece of length at least
$\mu N' - \nu - \beta$ and height $u[n]/\sqrt{N'}$. The remaining at
most $\nu+\beta$ symbols in the block 
\begin{equation*}
    [(n-1)\mu N' + 1 \scs n\mu N']
\end{equation*}
are each equal to $-u[n]/\sqrt{N'}.$ Hence, we have the following lower
bound on the signal in $u_k[n]$ (the term inside parentheses in
\eqref{eq:op_IDC}):
\begin{align}
    \label{eq:lb1}
    \sqrt{\frac{g}{\mu N'}}\sum_{\ell = (n-1)\mu N' + 1}^{n\mu N'}\tilde{x}_k[\ell] 
    & \geq \sqrt{\frac{g}{\mu N'}} (\mu N' -\nu -\beta)\frac{u[n]}{\sqrt{N'}} 
    + \sqrt{\frac{g}{\mu N'}}(\nu +\beta) \frac{-u[n]}{\sqrt{N'}}\nonumber\\
    & = \sqrt{\mu g}\biggl(1-\frac{2\nu +2\beta}{\mu N'}\biggr) u[n]\nonumber\\
    & = \sqrt{\mu g}\biggl(1-2\delta_N\biggr) \sqrt{P_1},
\end{align}
where 
\begin{equation}
    \label{eq:deltadef}
    \delta_N 
    \defeq \frac{\nu +\beta}{\mu N'}
    = \frac{N^{7/2}+N^{3}}{\mu N^4} 
    \rightarrow 0
\end{equation}
as $N\to\infty$ using the definitions of $\nu$ and $\beta$ in
\eqref{eq:nudef} and \eqref{eq:betadef}.

We also have the simple upper bound 
\begin{equation}
    \label{eq:ub1} 
    \sqrt{\frac{g}{\mu N'}}\sum_{\ell = (n-1)\mu N' + 1}^{n\mu N'}\tilde{x}_k[\ell] 
    \leq \sqrt{\mu g}\cdot \sqrt{P_1}.
\end{equation}
obtained by assuming that there is no inter-symbol interference.

Similarly, since $u[n-1] = u[n+1]=-u[n]=-\sqrt{P_1}$, we can derive the
analogous bounds
\begin{align}
    \label{eq:ub2}
    \sqrt{\frac{g}{\mu N'}}\sum_{\ell = (n-2)\mu N' + 1}^{(n-1)\mu N'}\tilde{x}_k[\ell] 
    & \geq -\sqrt{\mu g}\cdot \sqrt{P_1}, \\ 
    \label{eq:ub3}
    \sqrt{\frac{g}{\mu N'}}\sum_{\ell = n\mu N' + 1}^{(n+1)\mu N'} \tilde{x}_k[\ell] 
    & \geq -\sqrt{\mu g}\cdot \sqrt{P_1}
\end{align}
for the signal in $u_k[n-1]$ and $u_k[n+1]$ (the terms inside
parentheses in \eqref{eq:op_IDC2} and \eqref{eq:op_IDC3}, respectively).

\subsubsection*{Second Hop}

For the analysis of the second hop, number the IDCs from the relays to
the destination as $1$ through $K$.  By the definition of $\hat{u}[n]$,
\begin{align}
    \label{eq:uhat}
    \hat{u}[n] 
    & = \frac{1}{\sqrt{\mu N'}}\sum_{\ell = (n-1)\mu N' + 1}^{n\mu N'} y[\ell]\nonumber\\
    & = \frac{1}{\sqrt{\mu N'}}\sum_{\ell = (n-1)\mu N' + 1}^{n\mu N'} 
    \biggl(\sqrt{h}\sum_{k=1}^K\tilde{v}_k[\ell] + z[\ell]\biggr)\nonumber\\
    & = \sqrt{\frac{h}{\mu N'}}\sum_{k=1}^K\sum_{\ell = (n-1)\mu N' + 1}^{n\mu N'} 
    \tilde{v}_k[\ell] + \hat{z}[n],
\end{align}
where similar to before
\begin{equation*}
    \hat{z}[n] \defeq 
    \frac{1}{\sqrt{\mu N'}}\sum_{\ell = (n-1)\mu N' + 1}^{n\mu N'} z[\ell]
    \sim\mc{N}(0,1).
\end{equation*}

Consider the $k$th IDC. The input to this IDC is a piecewise constant
sequence, where the $n$th piece, with value 
$\frac{\alpha}{\sqrt{N'}} u_k[n]$, aligns precisely with the block
\begin{equation*}
     [(n-1)N'+1\scs nN']. 
\end{equation*}
However, as before, the $n$th piece in the output of the IDC may not
align with the block 
\begin{equation*}
    [(n-1)\mu N' +1 \scs n\mu N']
\end{equation*}
due to the randomness of the IDC. Some of the output symbols
in this block could be equal to $\frac{\alpha}{\sqrt{N'}}u_k[n-1]$ or
$\frac{\alpha }{\sqrt{N'}}u_k[n+1]$ causing inter-symbol interference.
However, since we condition on $\mc{E}_{\text{IDC}}^c$, we know that
this block contains a piece of length at least $\mu N'-\nu-\beta$ and
height $\frac{\alpha}{\sqrt{N'}}\cdot u_k[n]$, and the remaining at most
$\nu +\beta$ symbols are equal to either $\frac{\alpha}{\sqrt{N'}}\cdot
u_k[n-1]$ or $\frac{\alpha}{\sqrt{N'}}\cdot u_k[n+1]$. 

For concreteness, assume that the number of symbols in this block that
take values $\frac{\alpha}{\sqrt{N'}}\cdot u_k[n-1]$,
$\frac{\alpha}{\sqrt{N'}}\cdot u_k[n]$ and
$\frac{\alpha}{\sqrt{N'}}\cdot u_k[n+1]$ are $L_{k,n-1}$, $L_{k,n}$, and
$L_{k,n+1}$, respectively. From the above discussion, note that 
\begin{align}
    \label{eq:llower}
    L_{k,n} & \geq \mu N'-\nu-\beta \\
    \shortintertext{and} \notag\\
    \label{eq:lupper}
    L_{k,n-1}+L_{k,n+1} & \leq \nu+\beta.
\end{align}
Continuing~\eqref{eq:uhat}, we thus have
\begin{align*}
    \hat{u}[n] 
    & = \sqrt{\frac{h}{\mu N'}}\sum_{k=1}^K\sum_{\ell = (n-1)\mu N' + 1}^{n\mu N'} 
    \tilde{v}_k[\ell] + \hat{z}[n] \\
    & = \sqrt{\frac{h}{\mu N'}}\sum_{k=1}^K \Bigl(L_{k,n} \frac{\alpha}{\sqrt{N'}} u_k[n] 
    + L_{k,n-1}\frac{\alpha}{\sqrt{N'}} u_k[n-1]
    + L_{k,n+1}\frac{\alpha}{\sqrt{N'}} u_k[n+1] \Bigr) + \hat{z}[n] \\
    & \geq \sqrt{\frac{h}{\mu N'}}\sum_{k=1}^K \Bigl(L_{k,n} \frac{\alpha}{\sqrt{N'}} 
    \bigl((1-2\delta_N)\sqrt{\mu g P_1}+ \hat{z}_k[n]\bigr) \\
    & \qquad\qquad\qquad {} + L_{k,n-1}\frac{\alpha}{\sqrt{N'}}
    \bigl(-\sqrt{\mu g P_1}+\hat{z}_k[n-1]\bigr)+ L_{k,n+1}\frac{\alpha}{\sqrt{N'}}
    \bigl(-\sqrt{\mu g P_1}+\hat{z}_{k}[n+1]\bigr) \Bigr) + \hat{z}[n],
\end{align*}
where we have used \eqref{eq:op_IDC}, \eqref{eq:op_IDC2},
\eqref{eq:op_IDC3}, \eqref{eq:lb1}, \eqref{eq:ub2}, and \eqref{eq:ub3}.

Simplifying and using \eqref{eq:deltadef}, \eqref{eq:llower}, and
\eqref{eq:lupper} yields
\begin{align*}
    \hat{u}[n] 
    & \geq K\alpha\mu\sqrt{gh}\left(1-4\delta_N\right)\sqrt{P_1} \\
    & \quad {} + \sqrt{\frac{h}{\mu N'}}\cdot\frac{\alpha}{\sqrt{N'}}
    \sum_{k=1}^K \bigl(L_{k,n-1}\hat{z}_k[n-1] + L_{k,n}\hat{z}_k[n]
    + L_{k,n+1}\hat{z}_k[n+1]\bigr) + \hat{z}[n].
\end{align*}

The second term in the last expression constitutes the end-to-end noise.
It is clear that it is normal with zero mean. The term inside
parentheses is independent of $\hat{z}[n]\sim\mc{N}(0,1)$. We can upper
bound the variance of the term in parentheses by assuming $L_{k,n} = \mu
N'$ and $L_{k,n-1}=L_{k,n+1}=0$ for all $k$. The total noise power is
then upper bounded by
\begin{align*}
    \E\Biggl(\biggl(\sqrt{\frac{h}{\mu N'}}\cdot\frac{\alpha}{\sqrt{N'}}\sum_{k=1}^K
    \mu N' \hat{z}_k[n] + \hat{z}[n]\biggr)^2\Biggr)
    & = \frac{h}{\mu N'}\cdot\frac{\alpha^2}{N'}K(\mu N')^2+1 \\
    & = K\alpha^2\mu h + 1.
\end{align*}

Note that the noise in the received signal $(\hat{u}[n])_{n=1}^N$ is not
i.i.d. across $n$ due to the leakage of the first-hop noise from
neighboring pieces, i.e., $\hat{z}_k[n]$ appears in the noise
expressions for blocks $n-1$, $n$, and $n+1$. However, we can ignore
this fact and assume that the noise is independent across blocks,
because by interleaving codewords of the outer code such that successive
symbols of a codeword are sufficiently apart, they can be made to
experience independent noise. This allows us to treat the channel as an
AWGN channel. A scheme that makes use of the dependence of noise across
blocks can potentially achieve higher rates, but for simplicity of
analysis  we restrict attention to schemes that interleave codewords and
treat noise as independent across blocks.

The end-to-end SNR of the effective AWGN channel is at least
\begin{align*}
    \frac{\bigl(K\alpha\mu\sqrt{ghP_1}(1-4\delta_N)\bigr)^2}{K\alpha^2\mu h + 1}
    \geq \frac{K^2\mu^2ghP_1P_2}{1 + \mu gP_1 + K\mu hP_2}(1-8\delta_N),
\end{align*}
where we have used the definition of $\alpha = \sqrt{P_2/(1+\mu gP_1)}$
in~\eqref{eq:alpha}. Since $\delta_N \rightarrow 0$ as $N\to\infty$ by
\eqref{eq:deltadef}, we have that the SNR is lower bounded by $\SNR_{lb}$, where $\SNR_{lb}$ is given by
\begin{equation}\label{eq:SNR_lower_bound}
    \SNR_{lb} \triangleq \frac{K^2\mu^2P_1P_2gh}{1 + \mu gP_1 +
    K\mu hP_2}.
\end{equation}

Let $C_{\text{binary}}(\SNR)$ denote the largest achievable rate with binary antipodal constellation of the AWGN
channel as a function of SNR, and $C(\SNR)=\frac{1}{2}\log(1+\SNR)$ denote the capacity of the AWGN channel. Since the achievability scheme employs a binary antipodal constellation, the largest rate that can be achieved is given by $C_{\text{binary}}(\SNR)$ which, by \eqref{eq:SNR_lower_bound}, is lower-bounded by
$$C_{\text{binary}}\left(\SNR_{lb}\right),$$
which we will show later can be further lower bounded by
\begin{equation}\label{eq:gamma}\gamma C\left(\SNR_{lb}\right) = \frac{\gamma}{2}\log\left(1+\SNR_{lb}\right),\end{equation}
for all $\SNR_{lb}$ of interest, where $\gamma$ is a constant.

Using this, the achievable scheme we have
considered can transmit at a rate arbitrarily close to
\begin{equation*}
    \frac{\gamma}{2}\log\left(1+\SNR_{lb}
    \right)
\end{equation*}
bits per channel use with probability of error
\begin{equation}
    \label{eq:error2}
    \Pp(\mc{E} \mid \mc{E}_{\text{IDC}}^c)
    < \varepsilon/2
\end{equation}
for sufficiently large $N$.

\subsection*{Rate per Unit Energy}

Substituting~\eqref{eq:error1} and~\eqref{eq:error2}
into~\eqref{eq:error} shows that the overall probability of error is at
most $\Pp\left(\mc{E}\right)<\varepsilon$ for $N$ large enough.
Since the total energy used during transmission is no more than
$NP_1+KNP_2$ by~\eqref{eq:energy1} and~\eqref{eq:energy2}, the achieved
bits per unit energy is
\begin{IEEEeqnarray*}{Cl}
\geq & \frac{NC_{\text{binary}}(\SNR)}{N(P_1+KP_2)}\\ 
\geq & \frac{C_{\text{binary}}(\SNR_{lb})}{P_1+KP_2}\\
\geq & \frac{\gamma C(\SNR_{lb})}{P_1+KP_2}\\
 = &  \frac{\frac{\gamma}{2}\log\left(1+\frac{K^2\mu^2ghP_1P_2}{1 + \mu gP_1 + K\mu hP_2}\right)}{P_1+KP_2}.
\end{IEEEeqnarray*}

We now choose $P_1$ and $P_2$ depending on the regime of $(g,h,K)$. We
consider the cases $h<g/K$, $g/K \leq h < Kg$, and $h \geq Kg$
separately.
\begin{itemize}
    \item If $h < g/K$, we choose
        \begin{align*}
            P_1 & = \frac{1}{\mu g}, \\
            P_2 & = \frac{1}{\mu K^2 h}.
        \end{align*}
        Then the achievable bits per energy are
        \begin{align}
            \label{eq:case1}
            \frac{\frac{\gamma}{2}\log\bigl(1+\frac{1}{1 + 1 + 1/K}\bigr)}{1/(\mu g)+1/(\mu Kh)} 
            \geq \frac{\gamma \mu \log(4/3)}{4} Kh,
        \end{align}
        where we have used $h < g/K$.
    \item If $g/K \leq h < Kg$, we choose
        \begin{align*}
            P_1 & = \frac{1}{\mu \sqrt{Kgh}}, \\
            P_2 & = \frac{1}{\mu \sqrt{K^3gh}}.
        \end{align*}
        Then the achievable bits per energy are
        \begin{align}
            \label{eq:case2}
            \frac{\frac{\gamma}{2}\log\Bigl(1
            +\frac{1}{1 + \sqrt{g/(Kh)}+\sqrt{h/(Kg)}}\Bigr)}
            {1/(\mu \sqrt{Kgh})+1/(\mu \sqrt{Kgh})} 
            & = \frac{\gamma\mu\sqrt{Kgh}}{4}\log\biggl(1+\frac{1}{1 + 
            \sqrt{g/(Kh)} + \sqrt{h/(Kg)} }\biggr)\nonumber\\
            & \stackrel{(a)}{\geq} \frac{\gamma \mu \log(4/3)}{4}\sqrt{Kgh},
        \end{align}
        where we have used $1/\sqrt{K} \leq \sqrt{h/g} < \sqrt{K}$.
    \item If $h \geq Kg$, we choose 
        \begin{align*}
            P_1 & = \frac{1}{\mu Kg}, \\
            P_2 & = \frac{1}{\mu Kh}.
        \end{align*}
        Then the achievable bits per energy are
        \begin{align}
            \label{eq:case3}
            \frac{\frac{\gamma}{2}\log\bigl(1+\frac{1}{1 + 1/K + 1}\bigr)}
            {1/(\mu Kg)+1/(\mu h)} 
            & \geq \frac{\gamma \mu \log(4/3)}{4}Kg,
        \end{align}
        where we have used $h \geq Kg$.
\end{itemize}

Hence from \eqref{eq:case1}, \eqref{eq:case2} and \eqref{eq:case3}, we
conclude that the rate per unit energy achieved by the proposed scheme
is at least
\begin{align*}
    & \hspace{-1cm}\frac{\gamma \mu \log(4/3)}{4}\cdot
    \begin{cases}
        Kh, & \text{ if } h < g/K,\\
        \sqrt{Kgh}, & \text{ if } g/K \leq h < Kg,\\
        Kg, & \text{ if } h \geq Kg.
    \end{cases} \\
    & \vspace{1cm} \geq \frac{\gamma \mu \log(4/3)}{4}\cdot \min\bigl\{Kg,\sqrt{Kgh},Kh\bigr\}.
\end{align*}

It remains only to find the constant $\gamma$ in \eqref{eq:gamma}. We would like to choose $\gamma$ as large as possible such that ${C_{\text{binary}}(\SNR_{lb})\geq \gamma C(\SNR_{lb})}$ holds for all $\SNR_{lb}$ of interest. In each of the three
regimes of $(g,h,K)$, the powers $P_1$ and $P_2$ are chosen such that the value of $\SNR_{lb}$ is at most $\frac{1}{1+1+1/K}$, which is contained in $[1/3, 1/2]$.  Hence, $\gamma$ can be chosen as
\begin{align*}
    \gamma & = \inf_{1/3 \leq \SNR_{lb} \leq 1/2}
    \frac{C_{\text{binary}}(\SNR_{lb})}{C(\SNR_{lb})} \\
    & = \frac{C_{\text{binary}}(1/2)}{C(1/2)} \\
    & = \frac{0.29048}{0.29248} \\
    & \geq 0.99,
\end{align*}
 where the numerical values can be
computed as described for example in~\cite[pp.~435--437]{proakis12}.

Hence the rate per unit energy achieved by the proposed scheme
is at least
\begin{equation*}
    \frac{\gamma \mu \log(4/3)}{4}\cdot \min\bigl\{Kg,\sqrt{Kgh},Kh\bigr\} 
    \geq \frac{\mu}{10} \min\bigl\{Kg,\sqrt{Kgh},Kh\bigr\}.
\end{equation*}
This concludes the proof of Theorem~\ref{thm:unsync}.\hfill\IEEEQED

\subsection{Generalization to unequal drifts}\label{subsec:unequal_mu}
If the $k$th IDC in the first hop has mean $\tilde{\mu}_{1,k}$ and the $k$th IDC in the second hop has mean $\tilde{\mu}_{2,k}$, for $k\in\{1,2,\dots , K\}$, then in place of \eqref{eq:alpha}, the $k$th relay chooses a scaling factor $\alpha_k$ which is
$$\alpha_k = \sqrt{\frac{P_{2k}}{1+\tilde{\mu}_{1,k}gP_1}},$$
where $P_{2k}$ is the transmit power at the $k$th relay. Instead of using a repetition code of the same length at all the relays, the length of the repetition codes used by the $k$th relay is chosen to be $N_k'$ such that $\tilde{\mu}_{2,1}N_1'=\tilde{\mu}_{2,2}N_2'=\dots \tilde{\mu}_{2K}N_K',$ and the $N_k'$'s are sufficiently large, say $N_1'=N^4.$ Then, proceeding as in the previous subsection with a few minor changes, we end up with the following lower bound on the end-to-end SNR of the effective AWGN channel, a more general form than \eqref{eq:SNR_lower_bound}:
$$ \frac{\left(\sum_{k=1}^K \alpha_k\sqrt{\tilde{\mu}_{1,k}\tilde{\mu}_{2,k}gh} \right)^2 P_1}{1 + h\sum_{k=1}^K \tilde{\mu}_{2,k}\alpha_k^2 }.$$

Depending on the regime of $(g,h,K)$, we make the following choices, which yield after some simplification, the claim in \eqref{eq:unequal_mu}.
\begin{itemize}
    \item If $h < g/K$, we choose
        \begin{align*}
            P_1 & = \frac{1}{\left(\min_{1\leq k\leq K}\tilde{\mu}_{1,k}\right) g},\\
            P_{2k} & = \frac{1}{\tilde{\mu}_{2,k} K^2 h}, \;\; 1\leq k\leq K.
        \end{align*}
    \item If $g/K \leq h < Kg$, we choose
        \begin{align*}
            P_1 & = \frac{1}{\left(\min_{1\leq k\leq K}\tilde{\mu}_{1,k} \right)\sqrt{Kgh}}, \\
            P_{2k} & = \frac{1}{\tilde{\mu}_{2,k} \sqrt{K^3gh}},\;\; 1\leq k\leq K.
        \end{align*}   
    \item If $h \geq Kg$, we choose 
        \begin{align*}
            P_1 & = \frac{1}{\left(\min_{1\leq k\leq K}\tilde{\mu}_{1,k}\right) Kg}, \\
            P_{2k} & = \frac{1}{\tilde{\mu}_{2,k} Kh},\;\; 1\leq k\leq K.
        \end{align*}        
\end{itemize}

\bibliographystyle{IEEEtran}
\bibliography{journal_abbr,sync}

\begin{thebibliography}{10}
\providecommand{\url}[1]{#1}
\csname url@rmstyle\endcsname
\providecommand{\newblock}{\relax}
\providecommand{\bibinfo}[2]{#2}
\providecommand\BIBentrySTDinterwordspacing{\spaceskip=0pt\relax}
\providecommand\BIBentryALTinterwordstretchfactor{4}
\providecommand\BIBentryALTinterwordspacing{\spaceskip=\fontdimen2\font plus
\BIBentryALTinterwordstretchfactor\fontdimen3\font minus
  \fontdimen4\font\relax}
\providecommand\BIBforeignlanguage[2]{{%
\expandafter\ifx\csname l@#1\endcsname\relax
\typeout{** WARNING: IEEEtran.bst: No hyphenation pattern has been}%
\typeout{** loaded for the language `#1'. Using the pattern for}%
\typeout{** the default language instead.}%
\else
\language=\csname l@#1\endcsname
\fi
#2}}

\bibitem{huang13}
Y.-C. Huang, U.~Niesen, and P.~Gupta, ``Energy-efficient communication in the
  presence of synchronization errors,'' \emph{arXiv:1301.6589 [cs.IT]}, Jan.
  2013, submitted to IEEE Trans. Inf. Theory.

\bibitem{dobrushin67}
R.~L. Dobrushin, ``Shannon's theorems for channels with synchronization
  errors,'' \emph{Problems Inform. Transm.}, vol.~3, no.~4, pp. 11--26, 1967.

\bibitem{mitzenmacher09}
M.~Mitzenmacher, ``A survey of results for deletion channels and related
  synchronization channels,'' \emph{Prob. Surv.}, vol.~6, pp. 1--33, 2009.

\bibitem{kanoria10}
Y.~Kanoria and A.~Montanari, ``On the deletion channel with small deletion
  probability,'' in \emph{Proc. IEEE ISIT}, June 2010, pp. 1002--1006.

\bibitem{kalai10}
A.~Kalai, M.~Mitzenmacher, and M.~Sudan, ``Tight asymptotic bounds for the
  deletion channel with small deletion probabilities,'' in \emph{Proc. IEEE
  ISIT}, June 2010, pp. 997--1001.

\bibitem{venkataramanan11}
R.~Venkataramanan, S.~Tatikonda, and K.~Ramchandran, ``Achievable rates for
  channels with deletions and insertions,'' in \emph{Proc. IEEE ISIT}, July
  2011, pp. 346--350.

\bibitem{drmota12}
M.~Drmota, W.~Szpankowski, and K.~Viswanathan, ``Mutual information for a
  deletion channel,'' in \emph{Proc. IEEE ISIT}, July 2012, pp. 2561--2565.

\bibitem{iyengar11}
A.~R. Iyengar, P.~H. Siegel, and J.~K. Wolf, ``Modeling and information rates
  for synchronization error channels,'' in \emph{Proc. IEEE ISIT}, July 2011,
  pp. 380--384.

\bibitem{golay49}
M.~J.~E. Golay, ``Note on the theoretical efficiency of information reception
  using {PPM},'' \emph{Proc. IRE}, vol.~37, p. 1031, Sept. 1949.

\bibitem{verdu90}
S.~Verd{\'u}, ``On channel capacity per unit cost,'' \emph{IEEE Trans. Inf.
  Theory}, vol.~36, no.~5, pp. 1019--1030, Sept. 1990.

\bibitem{schein00}
B.~Schein and R.~Gallager, ``The {G}aussian parallel relay network,'' in
  \emph{Proc. IEEE ISIT}, June 2000, p.~22.

\bibitem{schein01}
B.~Schein, ``Distributed coordination in network information theory,'' Ph.D.
  dissertation, Massachusetts Institute of Technology, 2001.

\bibitem{niesen10b}
U.~Niesen and S.~Diggavi, ``The approximate capacity of the {G}aussian
  {$N$}-relay diamond network,'' \emph{IEEE Trans. Inf. Theory}, vol.~59,
  no.~2, pp. 845--859, Feb. 2013.

\bibitem{parvaresh13}
F.~Parvaresh and R.~Etkin, ``Using superposition codebooks and partial
  decode-and-forward in low-{SNR} parallel relay networks,'' \emph{IEEE Trans.
  Inf. Theory}, vol.~59, no.~3, pp. 1704--1723, Mar. 2013.

\bibitem{elgamal06}
A.~{El Gamal}, M.~Mohseni, and S.~Zahedi, ``Bounds on capacity and minimum
  energy-per-bit for {AWGN} relay channels,'' \emph{IEEE Trans. Inf. Theory},
  vol.~52, no.~4, pp. 1545--1561, Apr. 2006.

\bibitem{avestimehr07}
A.~S. Avestimehr and D.~N.~C. Tse, ``Outage capacity of the fading relay
  channel in the low-{SNR} regime,'' \emph{IEEE Trans. Inf. Theory}, vol.~53,
  no.~4, pp. 1401--1415, Apr. 2007.

\bibitem{massey72}
J.~Massey, ``Optimum frame synchronization,'' \emph{IEEE Trans. Commun.},
  vol.~20, no.~2, pp. 115--119, Apr. 1972.

\bibitem{verdu89}
S.~Verd{\'u}, ``The capacity region of the symbol-asynchronous {G}aussian
  multiple-access channel,'' \emph{IEEE Trans. Inf. Theory}, vol.~35, no.~4,
  pp. 733--751, July 1989.

\bibitem{cover81}
T.~Cover, R.~McEliece, and E.~Posner, ``Asynchronous multiple-access channel
  capacity,'' \emph{IEEE Trans. Inf. Theory}, vol.~27, no.~4, pp. 409--413,
  July 1981.

\bibitem{hui85}
J.~Hui and P.~Humblet, ``The capacity region of the totally asynchronous
  multiple-access channel,'' \emph{IEEE Trans. Inf. Theory}, vol.~31, no.~2,
  pp. 207--216, Mar. 1985.

\bibitem{chandar10}
V.~Chandar, A.~Tchamkerten, and D.~Tse, ``Asynchronous capacity per unit
  cost,'' in \emph{Proc. IEEE ISIT}, June 2010, pp. 280--284.

\bibitem{shomorony12}
I.~Shomorony, R.~Etkin, F.~Parvaresh, and S.~Avestimehr, ``Bounds on the
  minimum energy-per-bit for bursty traffic in diamond networks,'' in
  \emph{Proc. IEEE ISIT}, July 2012, pp. 801--805.

\bibitem{yeung09}
R.~Yeung, N.~Cai, S.-W. Ho, and A.~Wagner, ``Reliable communication in the
  absence of a common clock,'' \emph{IEEE Trans. Inf. Theory}, vol.~55, no.~2,
  pp. 700--712, Feb. 2009.

\bibitem{proakis12}
J.~G. Proakis, M.~Salehi, and G.~Bauch, \emph{Contemporary Communication
  Systems Using Matlab}, 3rd~ed.\hskip 1em plus 0.5em minus 0.4em\relax Cengage
  Learning, 2012.

\end{thebibliography}

\end{document}